%% file: reviewers-misinformation.tex
\documentclass[]{article}
\usepackage{authblk}

\usepackage{booktabs} 

\usepackage[utf8]{inputenc} 
\usepackage[T1]{fontenc}    
\usepackage{hyperref}       
\usepackage{url}            
\usepackage{booktabs}       
\usepackage{amsfonts}       
\usepackage{nicefrac}       
\usepackage{microtype}      
\usepackage{fullpage}
\usepackage{longtable}
\usepackage{adjustbox}
\usepackage{hyperref}
\usepackage{url}
\usepackage{graphicx}
\usepackage{amsmath}
\usepackage{bbm}
\usepackage{paralist}
\usepackage{enumitem}
\usepackage{mathtools}
\usepackage{subfig}
\usepackage{graphicx}
\usepackage{tabularx}
\usepackage{stackengine}
\usepackage{booktabs}
\usepackage{hhline}
\usepackage[multiple]{footmisc}
\usepackage{bm}
\usepackage{mathtools}
\usepackage{subfig}
\usepackage{color}
\usepackage{Definitions}
\usepackage[ruled,noline,norelsize]{algorithm2e}

\newcommand{\curb}{\textsc{Curb}\xspace}

\newcommand{\badge}[1]{\ifmmode\text{\texttt{#1}}\else\texttt{#1}\fi}
\newcommand{\xhdr}[1]{\vspace{1.2mm}\noindent{{\bf #1.}}}

\newcommand{\explain}[2]{\underset{\mathclap{\overset{\uparrow}{#2}}}{#1}}
\newcommand{\explainup}[2]{\overset{\mathclap{\underset{\downarrow}{#2}}}{#1}}
\newcommand*{\defeq}{\coloneqq}

\newcommand\blfootnote[1]{%
  \begingroup
  \renewcommand\thefootnote{}\footnote{#1}%
  \addtocounter{footnote}{-1}%
  \endgroup
}

\graphicspath{{./FIG/}}

\title{Leveraging the Crowd to Detect and Reduce \\ the Spread of Fake News and Misinformation}

\author[1]{Jooyeon Kim$^{*}$}
\author[2,3]{Behzad Tabibian}
\author[1]{Alice Oh}
\author[2]{Bernhard Sch\"{o}lkopf}
\author[3]{\mbox{Manuel Gomez-Rodriguez}}

\affil[1]{KAIST, jooyeon.kim@kaist.ac.kr, alice.oh@kaist.edu}
\affil[2]{MPI for Intelligent Systems Systems, btabibian@tue.mpg.de, bs@tue.mpg.de}
\affil[3]{MPI for Software Systems, manuelgr@mpi-sws.org}

\date{}

\begin{document}


\maketitle

\begin{abstract}
\input{000abstract}
\end{abstract}

\blfootnote{$^{*}$\scriptsize This work was done during Jooyeon Kim'{}s internship at the Max Planck
Institute for Software Systems.}

\section{Introduction} 
\label{sec:intro}
\input{010introduction}

\section{Problem formulation} 
\label{sec:formulation}
\input{020formulation}

\section{Proposed algorithm} 
\label{sec:algorithm}
\input{030algorithm}

\section{Experiments} 
\label{sec:experiments}
\input{040experiments}

\section{Conclusions} 
\label{sec:conclusions}
\input{050conclusions}

{
\bibliographystyle{abbrv}
\bibliography{refs}
}

\section*{Appendix}
\input{060appendix}

\end{document}

%% file: 000abstract.tex
Online social networking sites are experimenting with the following \emph{crowd-powered} procedure to reduce the spread of fake news and 
misinformation: 
whenever a user is exposed to a story through her feed, she can flag the story as misinformation and, if the story receives \emph{enough} flags, 
it is sent to a trusted third party for fact checking. If this party identifies the story as misinformation, it is marked as \emph{disputed}. 
However, given the uncertain number of exposures, the high cost of fact checking, and the trade-off between flags and exposures, the above 
mentioned procedure requires careful reasoning and smart algorithms which, to the best of our knowledge, do not exist to date.

In this paper, we first introduce a flexible representation of the above procedure using the framework of marked temporal point processes. 
Then, we develop a scalable online algorithm, \curb, to select \emph{which} stories to send for fact checking and \emph{when} to do so to efficiently 
reduce the spread of misinformation with provable guarantees.
In doing so, we need to solve a novel stochastic optimal control problem for stochastic differential equations with jumps, which is of independent 
interest.
Experiments on two real-world datasets gathered from Twitter and Weibo show that our algorithm may be able to effectively reduce the spread of 
fake news and misinformation. 

%% file: 010introduction.tex
In recent years, social media and online social networking sites have become a major disseminator of false facts, urban legends, fake news, 
or, more generally, misinformation.
In this context, there are growing concerns that misinformation on these platforms has fueled the emergence of a \emph{post-truth} society, 
where debate is perniciously framed by the repeated assertion of talking points to which factual rebuttals by the media or independent experts
are ignored.
For example, the post-truth label has been widely used to describe the presidential campaign of Donald Trump in the 2016 U.S. 
elections\footnote{\scriptsize \url{https://www.washingtonpost.com/posteverything/wp/2016/06/16/why-the-post-truth-political-era-might-be-around-for-a-while/}} and 
the Leave campaign in the 2016 Brexit referendum\footnote{\scriptsize \url{https://www.theguardian.com/commentisfree/2016/may/13/boris-johnson-donald-trump-post-truth-politician}},
whose outcomes have been then called into question.

In an effort to curb the spread of misinformation, major online social networking sites, such as Facebook, Twitter or Weibo, are (considering) 
resorting to the crowd\footnote{\scriptsize \url{https://newsroom.fb.com/news/2016/12/news-feed-fyi-addressing-hoaxes-and-fake-news/}}\footnote{\scriptsize \url{https://www.washingtonpost.com/news/the-switch/wp/2017/06/29/twitter-is-looking-for-ways-to-let-users-flag-fake-news/}}\footnote{\scriptsize \url{http://www.scmp.com/news/china/policies-politics/article/2055179/how-chinas-highly-censored-wechat-and-weibo-fight-fake}}.
In particular, they are experimenting with the following procedure to reduce the spread of misinformation:
whenever users are exposed to a story through their feeds, they have a choice to flag the story as misinformation and, if the story receives \emph{enough} 
flags, it is directed to a coalition of independent organizations\footnote{\scriptsize This coalition of organizations includes, among many others, 
Snopes~(\url{http://www.snopes.com}), FactCheck~(\url{http://www.factcheck.org}) or Politifact (\url{http://www.politifact.com}).}, signatories of Poynter'{}s International 
Fact Checking Code of Principles\footnote{\scriptsize \url{http://www.poynter.org/fact-checkers-code-of-principles/}}, for fact checking. 
If the fact checking organizations identify a story as misinformation, it gets flagged as disputed and may also appear lower in the users'{} feeds, reducing the 
number of people who are \emph{exposed} to misinformation.
In this context, online social networking sites are giving advice to its millions of users on how to spot misinformation 
online\footnote{\scriptsize \url{https://www.nytimes.com/2017/05/08/technology/uk-election-facebook-fake-news.html}}.
However, the above mentioned procedure requires careful reasoning and smart algorithms which, to the best of our knowledge, are nonexistent 
to date:

\vspace{2mm} \hspace{-4.7mm} \emph{--- Uncertain number of exposures:} the spread of information over social networking sites is a stochastic 
process, which may depend on, \eg, the information content, the users'{} influence and the network structure. Thus, the number of 
users exposed to different stories varies greatly and we need to consider probabilistic exposure models to capture this uncertainty.

\vspace{2mm} \hspace{-4.7mm} \emph{--- Fact checking is costly:} given the myriad of (fake) stories spreading in online social networking sites and the observation
that fact checking is a costly process, we can only expect (the reviewers from) the coalition of independent organizations to fact check a small percentage of 
the set of stories spreading over time.
Therefore, it is necessary to decide \emph{which} stories to fact check and \emph{when} to do so.

\vspace{2mm} \hspace{-4.7mm} \emph{--- Flags vs exposures:} the more users are exposed to a story before sending it for fact checking, the greater the confidence a story may be
misinformation, however, the higher the potential damage if it turns out to be misinformation. 
Thus, we need to find the optimal trade-off between misinformation evidence, by means of flagging data, and misinformation reduction, by means of preventing (unwarned) exposures 
to misinformation.

%
%
 
\xhdr{Our approach} To tackle the above challenges, we first introduce a novel representation of the above procedure using the framework of marked 
temporal point processes~\cite{AalBorGje08}. 
Then, we find \emph{which} stories to send for fact checking by solving a novel stochastic optimal control problem for 
SDEs with jumps~\cite{hanson2007applied}, which differs from the nascent literature on stochastic optimal control of social and 
information systems~\cite{redqueen17wsdm, wang2017variational, cheshire2017zarezade, wang2017icml} 
in two technical aspects:

\begin{itemize}
\item[I.] The control signal is a multidimensional survival process (\ie, a terminating temporal point process), which is defined by means of a set of conditional 
intensities (\ie, stories to fact check), while previous work has considered nonterminating temporal point processes as control signals.

\item[II.] The SDEs with jumps depend on random variables (\ie, flags) whose distributions depend on \emph{a priori} unknown parameters (\ie, flagging 
probability, which depends on whether a story is fake or legitimate). 
In this context, a posterior estimate of these parameters arises naturally in the formulation of the optimal control problem. 
In contrast, previous work did not consider application scenarios where posterior inference was required.
\end{itemize}
These technical aspects have implications beyond the problem of detecting and reducing misinformation since they establish an unexplored connection between
stochastic online optimal control, survival analysis and Bayesian inference.
Moreover, we find that the solution to the above optimal control pro\-blem is relatively simple and intuitive: the optimal intensity of fact checking is proportional to
a posterior estimate of the rate of misinformation.
As a consequence, we can develop a scalable online algorithm, \curb, to schedule \emph{which} stories to send for fact checking to efficiently reduce misinformation.
Finally, we experiment on two real-world datasets gathered from Twitter and Weibo and show that our algorithm is able to effectively reduce the spread of misinformation and
fake news. To facilitate research in this area, we release an open source implementation of our algorithm at \url{http://learning.mpi-sws.org/curb/}.


\subsection{Related work} 
The research areas most closely related to ours are on truth discovery and rumor detection.
In terms of truth discovery, most previous work discovers the truth by assessing the trustworthiness of the information sources and they do so using link-based 
measures~\cite{borodin2005link,gyongyi2004combating,10.1371/journal.pone.0128193}, information retrieval based measures~\cite{wu2007corroborating}, accuracy-based 
measures~\cite{dong2014data,dong2015knowledge,xiao2016towards,kumar2016disinformation,lumezanu2012bias}, content-based measures~\cite{adler2007content, zhao2015enquiring, tanushree2017cscw,wei2017learning,volkova2017separating,gupta2014tweetcred,Ruchansky2017deepfake}, 
graphical models~\cite{pasternack2013latent,yin2011semi,zhao2012probabilistic,zhao2012bayesian}, or survival analysis~\cite{reliability2017tabibian, lukasik2016hawkes}.
A recent line of work~\cite{li2015discovery,liu2011online,pal2012information,wang2014towards} also considers scenarios in which the truth may change 
over time.
In terms of rumor detection, previous work has typically relied on textual, network and temporal features to characterize and detect rumors~\cite{friggeri2014rumor, qazvinian2011rumor, kwon2017rumor, 
liu2015real, ma2016detecting, mendoza2010twitter, ratkiewicz2011truthy}.
However, our work differs from previous work in several key aspects:
(i) the \emph{crowd-powered} fact che\-cking procedure is, to the best of our knowledge, novel; 
(ii) we both detect and reduce the spread of misinformation, in contrast, previous work has focused on 
detection and has not designed interventional strategies considering the exposure dynamics; 
and, (iii) we develop a principled online algorithm by solving a stochastic optimal control of SDEs with 
jumps, which is able to adapt to changes in the rate of exposures and flags, while previous work has 
developed off-line algorithms.
Moreover, note that our methodology could be extended to leverage the above detection methods to further refine the posterior estimate of the rate of misinformation.

Learning from the crowd has been widely used in many different areas, \eg, web security~\cite{Chia2012security, Wang2012Sybil}, spam detection~\cite{zheleva2008trusting, chen2015TruSMS}, 
phishing detection~\cite{Moore2008phishing} and fake online account detection~\cite{Freeman2017fakeAccount}. Moreover, using expert knowledge, such as the one provided by fact checking 
organizations, has been used to improve the quality and reliability of crowd learning procedures~\cite{Liu2017expertValidation, Hung2015validateCrowd}.
However, to the best of our knowledge, the present work is the first that leverages both the crowd and expert knowledge in the context of detecting and preventing the spread of fake news
and misinformation.
%

Finally, there has been a paucity of work on stochastic optimal control of SDEs with jumps~\cite{redqueen17wsdm, wang2017variational, cheshire2017zarezade, wang2017icml}, however, previous work 
has focused on steering social activity, has not considered survival processes as control signals, and has not considered SDEs with jumps depending on random variables whose distributions depend on 
unknown parameters.

%% file: 020formulation.tex
To reduce the spread of misinformation, we optimize the follo\-wing fact checking procedure used by major online social networking sites.
Users can flag any story in their feed as misinformation and, if a story receives \emph{enough} flags, it is sent to a third party for fact checking.
If the third-party identifies a story as misinformation, it gets flagged as disputed and may also appear lower in the users'{} feeds.
%
%
In this procedure, since the third-party fact checking is costly, we need to decide \emph{which} stories to fact check and \emph{when} to do so---decide how many flags
are \emph{enough}. For ease of exposition, we assume that, if a story is sent for fact checking, the story is instantaneously verified---it is instantly revealed whether the story 
is fake or genuine\footnote{\scriptsize One could easily relax this assumption by modeling the delay between the time when a story is sent for fact checking and the
time when it is verified.}.

In this section, we first leverage the framework of marked temporal point processes~\cite{AalBorGje08} to model the above mentioned fact checking procedure,
starting from the data representation the model uses, then define and estimate the rate of misinformation, which we will use to decide what and when to fact check,
and finally state the fact checking scheduling problem.




\subsection{Data representation}
Given an online social networking site with a set of users $\Ucal$ and a set of unverified stories $\Scal$, we define two types of
user events:
\emph{exogenous} events, which correspond to the publication of stories by users on their own initiative, and
\emph{endogenous} events, which correspond to the resharing and/or flagging of stories by users who are exposed to
them through their feeds, similarly as in previous work~\cite{Farajtabar2014, de2016learning}.

Formally, we represent each exogenous event as a triplet
\begin{equation*}
p~~\defeq~~(~\explain{u}{\text{\vphantom{k}user}},\quad \explain{t}{\text{time}},\quad \explainup{s}{\text{story}}~),
\end{equation*}
which means that user $u \in \Ucal$ posted story $s \in \Scal$ at time $t$ at her own initiative. Moreover, we denote the history of exogenous
events for each story $s$ up to time $t$ by $\Hcal^{p}_s(t) = \bigcup_{i: t_i < t} \{ p_i \, | \, s_i = s \}$.

Similarly, we represent each endogenous event as a 5-tuple
\vspace{-1mm}
\begin{equation*}
e~~\defeq~~(~\explain{u}{\text{\vphantom{k}user}},\quad \explain{t}{\text{time}},\quad \explainup{s}{\text{story}}, \quad \explain{r}{\text{reshare}}, \quad \explainup{f}{\text{flag}}~),
\end{equation*}
which means that user $u \in \Ucal$ was exposed to story $s \in \Scal$ at time $t$ and decided (not) to reshare it, $r = 1$ ($r = 0$), and/or (not) to flag it, $f=1$ ($f=0$).
Then, we denote the history of endogenous events for story $s$ up to time $t$ by $\Hcal^{e}_s(t) = \bigcup_{i: t_i < t} \{ e_i \, | \, s_i = s \}$.
Finally, we gather the history of exogenous and endogenous events for story $s$ up to time $t$ by $\Hcal_s(t) = \Hcal^{p}_s(t) \bigcup \Hcal^{e}_s(t)$ and the
overall history of exogenous and endogenous events up to time $t$ by $\Hcal(t) = \bigcup_{s \in \Scal} \Hcal_s(t)$.


\subsection{Fact checking generative process}
We represent the times of the exogenous and endogenous events within the social networking site using two multidimensional counting processes, $\Nb^{p}(t)$ and $\Nb^{e}(t)$, in which the $s$-th dimension,
$N^{p}_s(t)$ and $N^{e}_s(t)$, count the number of exogenous and endogenous events for story $s$ up to time $t$.
Following the literature on temporal point processes~\cite{AalBorGje08}, we characterize these counting processes using their corres\-ponding intensities, \ie,
\begin{align*}
\mathbb{E}[d\Nb^p(t) | \Hcal (t)] &= \lambdab^p(t) dt, \\
\mathbb{E}[d\Nb^e(t) | \Hcal (t)] &= \lambdab^e(t) dt,
\end{align*}
where $d\Nb^p(t)$ and $d\Nb^e(t)$ denote the number of exogenous and endogenous events in the window $[t, t+dt)$ and $\lambdab^p(t) := [\lambda_s^p(t)]_{s \in \Scal}$ and
$\lambdab^e(t) := [\lambda_s^e(t)]_{s \in \Scal}$ denote the vector of intensities associated to all stories.
Every time a user is exposed to a story $s$, the binary indicators $r$ and $f$ are sampled from two Bernoulli distributions, $\PP(r=1) = r_s$ and $\PP(f=1) = f_s$, where $r_s$ and $f_s$ are
story specific parameters which may depend on many complex factors, \eg, content, source.

Moreover, we represent the times when stories are sent to a trusted third-party for fact checking using a multidimensional binary counting process, $\Mb(t)$, in which the $s$-th dimension, $M_s(t)$,
becomes one when story $s$ is sent for fact checking. Here, we characterize this process using a vector of intensities $\ub(t) := [u_s(t)]_{s \in \Scal}$ associated to all
stories, \ie,
%
$\mathbb{E}[d\Mb(t) | \mathcal{H}(t)] = \ub(t) \odot (1-\Mb(t))dt$,
%
where $\odot$ denotes pairwise product and the term $(1-\Mb(t))$ ensures each story is sent for
fact checking only once.
Whenever story $s$ is sent for fact checking, we assume it gets instantly verified---it gets revealed whether $s$ is fake or genuine.
Under this characterization, deciding \emph{which} stories to fact check and \emph{when} to do so becomes a problem of finding the vector of intensities $\ub(t)$
of fact checking events.

\vspace{-1mm}
\subsection{Endogenous and exogenous intensities}
Following previous work~\cite{Farajtabar2014, coevolve15nips, redqueen17wsdm}, we model exogenous events for each unverified story $s \in \Scal$ as:
%
\begin{equation} \label{eq:lambda-exogenous}
\lambda^{p}_s(t) := h_s(t) (1 - M_s(t)),
\end{equation}
where $h_s(t) \geq 0$ is a time-varying (differentiable) function and $(1-M_s(t))$
ensures the intensity of endogenous events becomes zero if the story is verified (\ie, fact checked).
%
%
For endogenous events, we consider the following form for the intensity functions:
%
%
\begin{align} \label{eq:lambda-endogenous}
\lambda^{e}_s(t) &= \Big [   \overbrace{ \int_{0}^{t}{ r_s(\tau) g(t - \tau) dN^e_{s}(\tau)} }^{ \text{\scriptsize Reshares} } + \overbrace { \int_{0}^{t}{ g(t - \tau) dN^p_{s}(\tau)} }^{\text{\scriptsize Posts}}  \Big ] \overbrace{(1 - M_s(t))}^{\text{\scriptsize{Fact check}}},
\end{align}
where the first term, with $r_s(\tau) \sim \text{Bernoulli}(r_s)$, models the exposures due to reshares of story $s$ by the previous users, the
second term models the exposures due to previous posts of story $s$ by users on their own initiative (\ie, exogenous events).
%
%
In the first two terms, $g_s(t)$ denotes an exponential triggering kernel, $g(t) := \gamma \exp(- \omega t) \mathbb{I} (t \geq 0)$, which models
the decay of \emph{influence} of previous reshares and posts over time.
By making the intensity dependent on the history of reshares and posts, it becomes a stochastic process by itself.

Given the above functional form for the intensity of endogenous events, the following alternative representation based on stochastic differential equations (SDEs) with jumps,
which can be derived using Ito'{}s calculus~\cite{hanson2007applied}, will be useful to design our algorithm: 
\begin{proposition}
\label{first_proposition}
Let $N^{e}_s(t)$ be a counting process with associated intensity $\lambda^{e}_s(t)$, given by Eq.~\ref{eq:lambda-endogenous}. The tuple $(N^{e}_s(t), \lambda^{e}_s(t))$ is a doubly stochastic Markov process whose
dynamics can be represented by the following SDE with jumps:
\begin{align}
  d\lambda^{e}_s(t) &= -  \omega \lambda^{e}_s(t) dt - \lambda^{e}_s(t) dM_s(t) + \gamma \left[ r_s(t) dN^{e}_{s}(t) + (1-M(t)) dN^{p}_{s}(t) \right]  \label{eq:dlambda-endogenous}
\end{align}
\end{proposition}
%

\vspace{-1mm}
\subsection{Estimated rate of misinformation}
If one cannot send all stories for fact checking, ideally, one may like to send only fake stories and favor those which, if not fact checked, would
reach the greatest number of users.
However, we cannot directly observe whether a story is fake (that is why we send it for fact checking!) and we do not know the total number of users
who will be exposed to the story if not fact checked.
Instead, for each story, we will leverage the users'{} flags and exposures to compute a running estimate of the rate of misinformation due to that story.
Then, we will find the optimal fact checking intensities $\ub(t)$ that minimize a nondecreasing function of the estimated misinformation rates
over time.


Given a story $s \in \Scal$, the number of users $N^{f}_s(t)$ who flagged the story up to time $t$ can be formally defined as
\begin{equation*} 
N^f_s(t) := \int^{t}_{0}{f_s(\tau) dN^{e}_s(\tau)},
\end{equation*}
where $N^{e}_s(t)$ counts the number of users exposed to story $s$ by time $t$, $f_s(\tau) \sim f_s$, and it readily follows that
$dN^f_{s}(t) = f_s(t)dN^{e}_s(t)$.
%
%
%
%
Then, we can estimate the average number of users $\bar{N}^{m}_s(t)$ exposed to misinformation by time $t$, conditioned on the
number of exposed users and flags, as
\begin{align*} 
\bar{N}^{m}_s(t) &:= p_{m | s, f=1} N^f_s(t) + p_{m | s, f=0} ( N^{e}_s(t)  - N^f_s(t) ) = p_{m | f=1} N^f_s(t) + p_{m | f=0} ( N^{e}_s(t)  - N^f_s(t) )
\end{align*}
where we assume that the probability that a story is misinformation given that a user did or did not flag it is equal for all stories,
\ie, $p_{m | s, f} = p_{m | f}$, and one can estimate $p_{m | f}$ from historical flag and exposure data about fact checked stories.
Here, the probability $p_{m | f}$ characterizes how \emph{good} the crowd is at spotting misinformation\footnote{\scriptsize One
could consider different false positive and false negative rates per user and per story, \eg, using text analysis or domain knowledge
about the website that published the story. However, for simplicity, we leave such extensions for future work.}.

Next, we can compute the differential $d\bar{N}^{m}_s(t)$ as
\begin{equation*} 
d\bar{N}^{m}_{s}(t) = \Big [  (p_{m | f=1} - p_{m | f=0})f_s(t) + p_{m | f=0}  \Big ] dN^{e}_s(t)
\end{equation*}
and define the rate of misinformation as
\begin{align}
\lambda^{m}_s(t) dt &= \EE[d\bar{N}^{m}_{s}(t)]
= \big [  (p_{m | f=1} - p_{m | f=0})  f_s + p_{m | f=0} \big ] \lambda^{e}_s(t) dt. \label{eq:lambdam}
\end{align}
%
%
Unfortunately, the flagging probability $f_s$ may depend on many complex factors, \eg, content, source, and is generally
unknown.
To overcome this challenge, we assume a Beta prior on $f_s$, \ie, $f_s \sim Beta(\alpha, \beta)$, and compute
instead a posterior estimate of the rate of misinformation, which leverages both the number of exposures
$N^{e}_s(t)$ and flags $N^{f}_{s}(t)$ by time $t$, \ie,
\begin{align}
\hat{\lambda}^{m}_s(t)dt &= \EE_{f_s(t), f_s}[d\bar{N}^{m}_{s}(t)]
=\big [  (p_{m | f=1} - p_{m | f=0}) \frac{\alpha +  N^f_s(t) }{\alpha +  \beta + N^{e}_s(t) }
+ p_{m | f=0} \big ] \lambda^{e}_s(t) dt, \label{eq:lambdam-posterior}
\end{align}
where we used the conjugacy between the Bernoulli and the Beta distributions, \ie,
\begin{equation*}
f_s | N^{f}_s, N^{e}_s \sim Beta \left(\alpha + N^f_s(t),  \beta + N^{e}_s(t) - N^{f}_s(t)\right).
\end{equation*}
%
%
Finally, given a set of stories $\Scal$, we denote the vector of posterior estimates of their associated rate of misinformation
as $\hat{\lambdab}^{m}(t) = [\hat{\lambda}^{m}_s(t)]_{s \in \Scal}$.

\vspace{-1mm}
\subsection{The fact checking scheduling problem}
Given a set of unverified stories $\Scal$, we aim to find the optimal fact checking intensities $\ub(t)$ that minimize the expected value of nondecreasing
convex loss function $\ell( \hat{\lambdab}^m(t), \bm{u}(t) )$ of the posterior estimates of the stories rates of misinformation $\hat{\lambdab}^m(t)$ and the fact
checking intensities $\ub(t)$ over a time window $(t_0, t_f]$, \ie,
%
%
\begin{align} \label{eq:problem}
& \underset{\ub(t_0, t_f]}{\text{minimize}} \quad \,\, \mathbb{E} \left[ \phi( \bm{\hat{\lambda}}^m(t_f)  ) + \int_{t_0}^{t_f}{  l ( \bm{\hat{\lambda}}^m(\tau), \bm{u}(\tau))d\tau  }  \right]   \nonumber \\
& \text{subject to} \quad \bm{u}(t) \geq 0 \quad \forall t \in (t_0, t_f],
\end{align}
where $\ub(t_0, t_f]$ denotes the fact checking intensities from $t_0$ to $t_f$, the expectation is taken over all possible realizations of the marked temporal point processes associated
to the endogenous and exogenous events of all stories from $t_0$ to $t_f$, and $\phi(\bm{\hat{\lambda}}^m(t_f))$ is an arbitrary penalty function.
Here, by considering a nondecreasing loss on both the posterior estimate of the rates of misinformation and fact checking intensities, we penalize high levels of misinformation and we limit
the number of stories that are sent for fact checking.

%% file: 030algorithm.tex
In this section, we find the optimal fact-checking intensities $\ub(t)$ that minimizes Eq.~\ref{eq:problem} from the perspective of stochastic optimal control of jump SDEs~\cite{hanson2007applied}.
To ease the exposition, we first derive a solution for one story, introduce an efficient algorithm that implements the solution, and then generalize the solution and the efficient algorithm to multiple
stories.

\vspace{-1mm}
\subsection{Optimizing for one story}
Given an unverified story $s$ with fact checking reviewing intensity $u_s(t) = u(t)$, exogenous intensity $\lambda^{p}_s(t) = \lambda^{p}(t)$, endogenous intensity $\lambda^{e}_s(t) = \lambda^{e}(t)$ and associated
counting processes $M_s(t) = M(t)$, $N^{p}_s(t) = N^{p}(t)$ and $N^{e}_s(t) = N^{e}(t)$, respectively, we can rewrite the fact checking scheduling problem defined in Eq.~\ref{eq:problem} as:
%
%
\begin{align} \label{eq:problem-one-story}
& \underset{u(t_0, t_f]}{\text{minimize}} \quad \,\, \EE \left[ \phi( \hat{\lambda}^m(t_f)  ) + \int_{t_0}^{t_f} { l ( \hat{\lambda}^m(\tau), u(\tau))}d\tau  \right]   \nonumber \\
& \text{subject to} \quad u(t) \geq 0 \quad \forall t \in (t_0, t_f],
\end{align}
where the dynamics of $N^{e}(t)$ are given by Eq.~\ref{eq:dlambda-endogenous}
%
%
and the posterior estimate of the misinformation rate is given by Eq.~\ref{eq:lambdam-posterior}. Note that we are now focusing on a single story rather than on multiple stories.

%

%

Next, we define an optimal cost-to-go function $J$ for the above problem, use Bellman'{}s principle of optimality to derive the associated Hamilton-Jacobi-Bellman (HJB) equation~\cite{bertsekas1995dynamic},
and exploit the structure of our problem to find a solution to the HJB equation. 
\begin{definition} \label{def:cost-to-go}
The optimal cost-to-go $J$ is defined as the minimum of the expected value of the cost of going from state $(M(t), N^{e}(t), N^f(t), N^{p}(t), \lambda^{e}(t))$ at time $t$ to the final state
at time $t_f$, \ie,
  \begin{equation} \label{eq:cost-to-go}
    J(M(t), N^{e}(t), N^f(t), N^{p}(t), \lambda^e(t), t) = \underset{u(t, t_f]}{\text{min}}{\EE  \left[ \phi( \hat{\lambda}^m(t_f)  ) + \int_{t}^{t_f}{  \ell ( \hat{\lambda}^m(\tau), u(\tau))d\tau  }  \right]  }. \nonumber
  \end{equation}
\end{definition}
Now, using the Markov property of the state variables $(M(t), N^{e}(t), N^f(t), N^{p}(t), \lambda^{e}(t))$, we can apply Bellman'{}s principle of optimality to the above definition to break the
problem into smaller subproblems and rewrite Eq.~\ref{eq:cost-to-go} as
%
\begin{equation}
\label{eq: HJBeq1}
0 =  \underset{u(t, t + dt]}{\text{min}} \left\{\EE \left[ dJ( M(t), N(t), N^f(t), N^{p}(t), \lambda^{e}(t), t ) \right] + \ell( \hat{\lambda}^m(t), u(t))dt  \right\}.
\end{equation}
%
%
Next, we differentiate $J$ with respect to time $t$, $M(t)$, $N^{e}(t)$, $N^f(t)$, $N^{p}(t)$ and $\lambda^e(t)$ using Lemma~\ref{lem:dJ} (in the Appendix) with $x(t) = \lambda^{e}(t)$, $y(t) = \lambda^p(t) = h(t)$, $z(t) = r(t)$, $w(t) = f(t))$ and $F = J$.
\begin{align*}
    dJ &= \big [     J  (   M, N^{e} + 1, N^f + 1, N^p, \lambda^{e} + \gamma ,t ) f(t) r(t) + J (   M, N^{e} + 1, N^f + 1 , N^p, \lambda^{e}, t ) f(t) (1 - r(t))    \nonumber \\
    &+  J (   M,  N^{e} + 1, N^f, N^p, \lambda^{e} + \gamma , t ) (1-f(t)) r(t)  +  J (   M, N^{e} + 1, N^f , N^p, \lambda^{e}, t  ) (1 -  f(t))(1 - r(t)) \nonumber \\
    &- J(   M,  N^{e}, N^f , N^p, \lambda^{e}, t  )   \big ] dN^{e}(t) + \big [ J (   M,  N^{e}, N^f , N^p+1, \lambda^{e} + \gamma, t  ) \nonumber \\
    & - J(  M,  N^{e}, N^f , N^p, \lambda^{e}, t  )    \big ]  dN^p(t)  - \big [J( M, N^{e}, N^f , N^p, \lambda^{e}, t  ) - J( M + 1, N^{e}, N^f, N^p, 0, t) \big ] dM(t) \\
    & + J_t -\omega \lambda^{e}(t) J_{\lambda^{e}},
\end{align*}
Then, using that $\EE[dN^{e}(t)] = \lambda^{e}(t)dt$, $\EE[dN^p(t)] = h(t) dt$,
$\EE[dM(t)] = (1-M(t)) u(t)dt$, $\EE[r(t)] = r_s = r$,
and $\EE[f(t)] = \frac{\alpha + N^f(t)}{\alpha + \beta + N^{e}(t)}$, the HJB equation follows:
\begin{align}
    0 &= \big [     J  (   M, N^{e} + 1, N^f + 1, N^p, \lambda^{e} + \gamma ,t ) \tfrac{\alpha + N^f(t)}{\alpha + \beta + N(t)} r  + J (   M, N^{e} + 1, N^f + 1 , N^p, \lambda^{e}, t ) \tfrac{\alpha + N^f(t)}{\alpha + \beta + N(t)} (1 - r)    \nonumber \\
    &+  J (   M,  N^{e} + 1, N^f, N^p, \lambda^{e} + \gamma , t ) (1 - \tfrac{\alpha + N^f(t)}{\alpha + \beta + N(t)}  ) r +  J (   M, N^{e} + 1, N^f , N^p, \lambda^{e}, t  ) (1 -  \tfrac{\alpha + N^f(t)}{\alpha + \beta + N(t)}  )(1 - r) \nonumber \\
    &- J(   M,  N^{e}, N^f , N^p, \lambda^{e}, t  )   \big ] \lambda^{e}(t) + \big [ J (   M,  N^{e}, N^f , N^p+1, \lambda^{e} + \gamma, t  ) \nonumber \\
    & - J(  M,  N^{e}, N^f , N^p, \lambda^{e}, t  )    \big ] (1-M(t)) h(t)  + \underset{u(t, t_f]}{\text{min}} \Big\{ l(\hat{\lambda}^m(t), u(t)) + J_t -\omega \lambda^{e}(t) J_{\lambda^{e}} \nonumber \\ 
    & - \big [J( M, N^{e}, N^f , N^p, \lambda^{e}, t  ) - J( M + 1, N^{e}, N^f, N^p, 0, t) \big ] (1-M(t)) u(t) \Big \}, \label{eq:HJBeq2}
  \end{align}
%
with $J( M(t_f), N^{e}(t_f), N^f(t_f) , N^p(t_f), \lambda^{e}(t_f), t_f) = \phi( \hat{\lambda}^e(t_f) ) $ as the terminal condition.

To solve the above HJB equation, we need to define the penalty function $\phi$ and the loss function $\ell$. Following the literature on the stochastic optimal control~\cite{hanson2007applied},
we consider the following quadratic forms, which penalize high levels of misinformation and limit the number of stories that are sent for fact checking:
\begin{align*}
\phi(\hat{\lambda}^m(t_f)) &= \frac{1}{2}{ (\hat{\lambda}^{m}(t_f))^2} \\
\ell(\hat{\lambda}^m(t), u(t)) &= \frac{1}{2}{ (\hat{\lambda}^{m}(t))^2 } + \frac{1}{2}{ q u^{2} (t) },
\end{align*}
where $q$ is a tunable parameter that accounts for the trade-off between the number of stories sent for fact checking and the spread of misinformation in a social network.
%
%
With the specified loss function, we take the derivative with respect to $u(t)$ in Eq.~\ref{eq:HJBeq2} and uncover the relationship between the optimal fact checking intensity
and the optimal cost $J$:
\begin{align} \label{eq:u-sol}
u^{*}(t) &= q^{-1} (1-M(t)) \big [ J( M(t), N^{e}(t), N^f(t) , N^p(t), \lambda^{e}(t), t) - J( M(t) + 1, N^{e}(t), N^f(t) , N^p(t), 0, t) \big ].
\end{align}

Then, we plug in the above expression in Eq.~\ref{eq:HJBeq2} and find a solution to the resulting nonlinear differential equation using
the following Lemma:
\begin{lemma} \label{lem:optimal-J}
The optimal cost-to-go $J$ that satisfies the HJB equation, defined by Eq.~\ref{eq:HJBeq2}, is given by:
\begin{multline} \label{eq:cost-to-go-sol}
J(M(t), N^{e}(t), N^f(t), N^p(t), \lambda^e(t), t) = q^{\frac{1}{2}} \big [  (p_{m | f=1} - p_{m | f=0} ) \tfrac{\alpha + N^f(t)}{\alpha + \beta + N(t)} +  p_{m | f=0} \big ] \big [ \lambda^e(t) \\
 - \gamma  N^p(t) - (\gamma r - \omega) (\alpha + \beta + N^e(t))  \big ].
\end{multline}
%
\end{lemma}
\begin{xsketch}
We verify the above cost-to-go $J$ satisfies the HJB equation, given by Eq.~\ref{eq:HJBeq2}. To do so, we use that $J_t = 0$
and
\begin{equation*}
J_{\lambda^e} = q^{\frac{1}{2}} \big [  (p_{m | f=1} - p_{m | f=0} ) \tfrac{\alpha + N^f(t)}{\alpha + \beta + N(t)} +  p_{m | f=0} \big ].
\end{equation*}
Here, note that the HJB equation needs to be satisfied only for $M(t) \in \{0, 1\}$, which are the only feasible values for the counting
process associated to the fact checking event.
\end{xsketch}
\vspace{3mm}

Next, we use the above Lemma to recover the optimal fact checking intensity $u^{*}(t)$:
%
%
%
%
%
%
\begin{theorem}
Given a story $s$, the optimal fact checking intensity for the fact checking scheduling problem, defined by Eq.~\ref{eq:problem-one-story}, under quadratic loss
is given by:
\begin{equation}
\label{eq:u-optimal}
u^{*}(t) = q^{- \frac{1}{2}} (1-M(t)) \Big [ p_{m | f=0} + (p_{m | f=1} - p_{m | f=0} ) \left( \frac{\alpha + N^f(t)}{\alpha + \beta + N^{e}(t)} \right) \Big ] \lambda^{e}(t).
\end{equation}
\end{theorem}
The above result reveals a linear relationship between the optimal fact checking intensity and the endogenous (\ie, exposure) intensity $\lambda^{e}(t)$. Moreover,
the corresponding coefficient depends on the number of user exposures $N^{e}(t)$ and the number of flags $N^f(t)$ and can increase and decrease
over time.
Remarkably, the optimal intensity does not depend on the exogenous intensity $\lambda^p(t)$, which is generally unknown.

Finally, we derive an efficient procedure to sample from the above intensity, summarized in Algorithm~\ref{alg:curb}, which we name \curb.
Within the algorithm, $Next()$ returns the next endogenous (\ie, exposure) event, which occurs at time $t'{}$ with resharing and flagging indicators $r$ and $f$, respectively,
$Update(N^{e}(t), N^{f}(t), \lambda^{e}(t))$ updates $u(t)$ using Eq.~\ref{eq:u-optimal}, and $Sample(\tau, u(t))$ samples from an intensity $u(t)$ starting at time
$\tau$ using a standard thinning algorithm~\cite{lewis1979simulation}.
%
%
%
If the fact checking intensity increases, by $r = 1$ or $f = 1$, we apply the superposition theorem~\cite{kingman1993poisson}, \ie, we sample a new candidate fact
checking time $\kappa$ and we take the minimum between $\kappa$ and the previous candidate fact checking time.
If the fact checking intensity decreases, by $f = 0$, we first use a thinning argument to decide whether to keep the current candidate fact checking time $\tau$ or we
sample a new candidate fact checking time, greater than $\tau$.
This sampling procedure is very efficient since it only needs to sample $O(N^{e}(t_f))$ times from an inhomogeneous Poisson process.
Finally, note that if we gather more and more evidence that a story $s$ is not misinformation ($f=0$ for many exposures), the candidate fact checking time $\tau$ will
eventually become greater than $t_f$ and the story will never be sent for fact checking.
%
%
\SetKwInOut{Initialization}{Initialization}
\begin{algorithm}[t]
\DontPrintSemicolon 
\KwIn{Parameters $q$, $\alpha$, $\beta$, $p_{m|f=1}$, $p_{m|f=0}$, $t_f$}
\Initialization{$N(t) \gets 0;$ $N^f(t) \gets 0;$ $\lambda^e(t) \gets 0;$ \\ \hspace{0.5mm} $Update(N^e(t), N^{f}(t), \lambda^{e}(t))$}
\KwOut{Fact checking time $\tau$}
$\tau \gets t_f$\;
$\left(t'{}, r, f\right) \gets Next()$\;
\While{$t'{} < \tau$} {
    $u_{0}(t) \gets u(t)$\;
    $N^{e}(t) \gets N^{e}(t) + 1;\quad N^f(t) \gets N^f(t) + f$\;
    $u(t) \gets Update(N^e(t), N^{f}(t), \lambda^{e}(t))$\;
    \If{ f = 0 }{
    	$x \gets Uniform(0,1)$\;
	\If{ $u(t) / u_0(t) < x$ }{
    		$\tau \gets Sample\big (\tau, u(t) \big )$\;
	}
    }
    \If{ r = 1 }{
        $\lambda^e(t) \gets \lambda^e(t) + g(t - t'{})$\;
        $u(t) \gets Update(N^e(t), N^{f}(t), \lambda^{e}(t))$\;
    }
    $\kappa = Sample\left(t',  \max(0, u(t) - u_{0}(t))\right)$\;
    $\tau \gets min (\tau, \kappa)$\;
    $\left(t'{}, r, f \right) \gets Next()$\;
}
\Return{$\tau$}\;
\caption{The \curb Algorithm} \label{alg:curb}
\end{algorithm}

\vspace{-1mm}
\subsection{Optimizing for multiple stories}
Given a set of unverified stories $\Scal$ with fact checking intensities $\bm{u}(t)$, exogenous intensities $\lambdab^{p}(t)$, endogenous intensities $\lambdab^{e}(t)$ and associated
counting processes $\Mb(t)$, $\Nb^{p}(t)$ and $\Nb^{e}(t)$, we can solve the fact checking scheduling problem defined by Eq.~\ref{eq:problem} similarly as in the case of a single
story.
In particular, consider the following quadratic form for the penalty function $\phi$ and loss function $\ell$:
\begin{align*}
\ell(  \hat{\lambdab}^m(t), \ub(t)) = \frac{1}{2}{   \sum_{s \in \Scal} (\hat{\lambda}^{m}_{s}(t))^2 } + \frac{1}{2}{ \sum_{s \in \Scal} q_s u_s^{2}(t) },
\end{align*}
where $\{ q_s \}_{s \in \Scal}$ are given parameters, which trade off the number of stories sent for fact checking and the spread of misinformation, and may favor fact checking some
stories over others.
Then, we can derive the optimal control signal intensity for each story, using the independence assumption between stories, proceeding similarly as in the case of a single story.
\begin{theorem}
Given a set of stories $\Scal$, the optimal fact checking intensity for each story $s \in \Scal$, which minimizes Eq.~\ref{eq:problem}, under quadratic loss and penalty functions is
given by:
\begin{equation}
u^{*}_s(t) = q_s^{- \frac{1}{2}} (1-M_s(t)) \Big [ p_{m | f = 0} +  (p_{m|f=1} - p_{m | f = 0} ) \left( \frac{\alpha + N^f_s(t)}{\alpha + \beta + N^{e}_s(t)} \right)  \Big ] \lambda^{e}_s(t).
\end{equation}
\end{theorem}
Finally, we can sample the fact checking times for all stories by running $|\Scal|$ instances of \curb, one per story, and the overall number of required samples
is $O(\sum_{s \in \Scal} N^{e}_s(t_f))$.
Moreover, note that the instances can be run in parallel and thus our algorithm is highly scalable.

%% file: 040experiments.tex
In this section, we evaluate our algorithm on data ga\-thered from two social networking sites, Twitter and Weibo, and
compare its performance with three baselines via two metrics.
\begin{figure}[t]
      \centering
      \subfloat[Twitter]{\includegraphics[width=.48\textwidth]{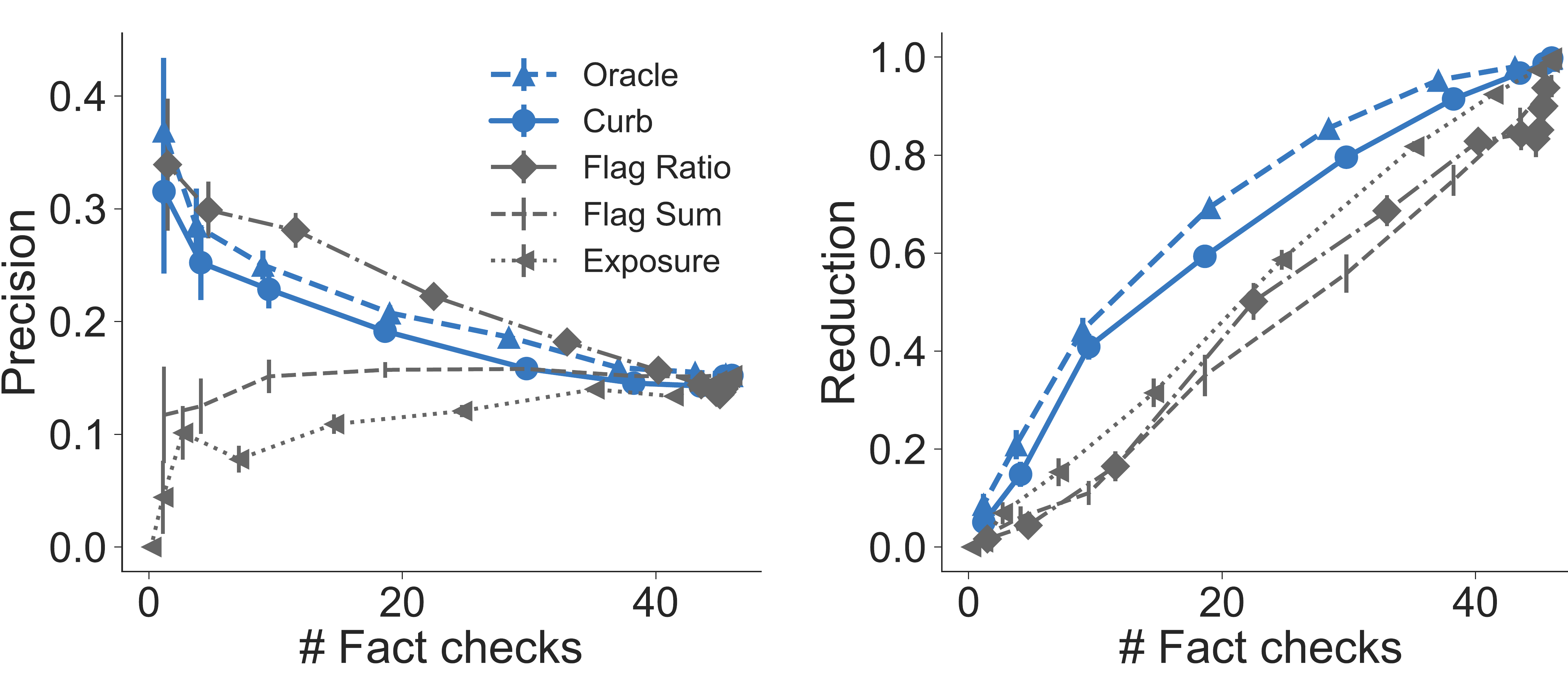}}
      \subfloat[Weibo]{\includegraphics[width=.48\textwidth]{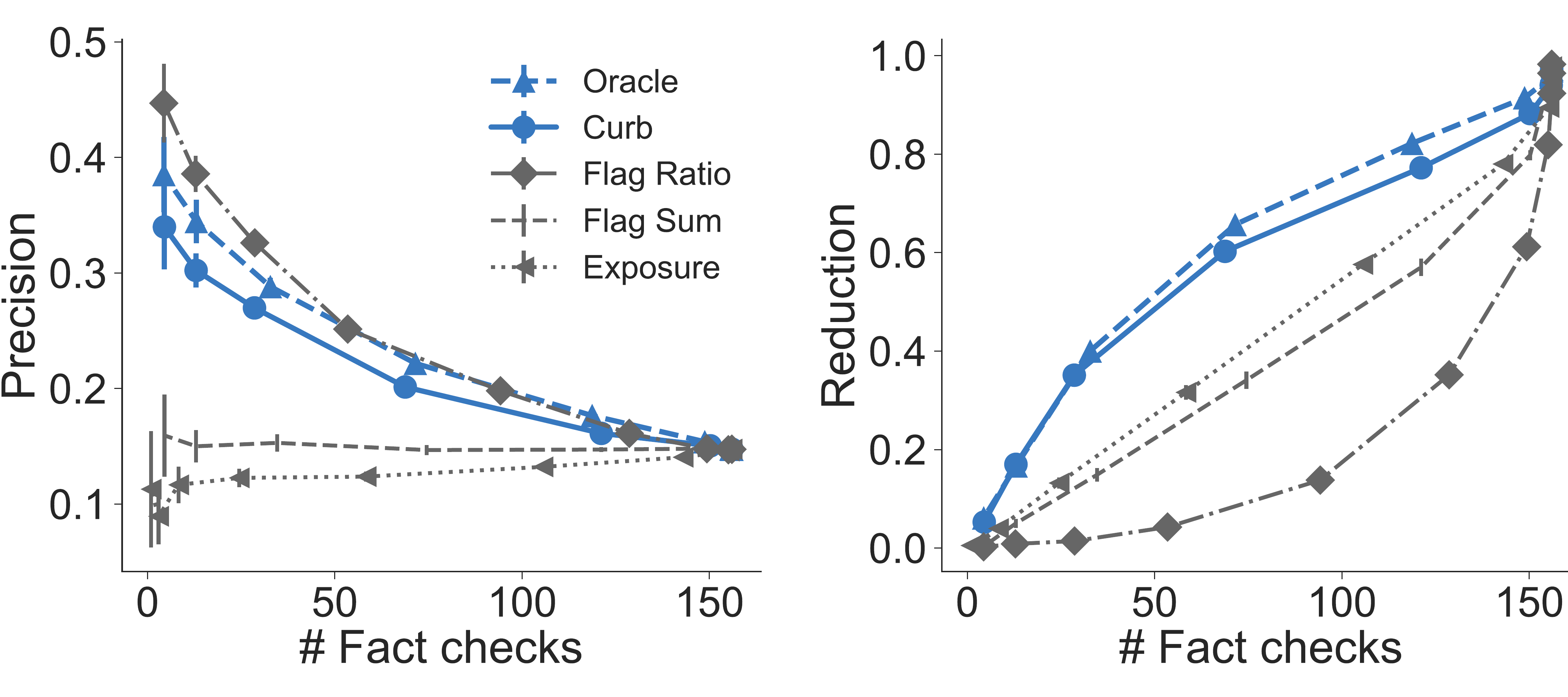}}
      \caption{Performance vs. number of fact checking events. 
      We measure performance in terms of misinformation reduction, which is the fraction of unverified exposures that fact checking prevented, and precision, which is the fraction of fact checked
      stories that are fake.
      The oracle and \curb achieve a comparable performance in both datasets, beating the threshold baseline by large margins.
      } \label{fig:tradeoff-number}
      \vspace{-3mm}
\end{figure}

\subsection{Dataset description and experimental setup}
We use data gathered from Twitter and Weibo as reported in previous work~\cite{kwon2017rumor,ma2016detecting}, which comprises posts and reshares for a variety of (manually annotated) genuine
and fake stories\footnote{\scriptsize In the Twitter and the Weibo datasets, stories were fact checked using \url{snopes.com} and the Sina community management center, \url{service.account.weibo.com}.}, respectively.
%
%
More specifically, the Twitter dataset contains $192{,}350$ posts and reshares from $117{,}824$ users for $111$ unique stories. The Weibo dataset contains $3{,}752{,}459$ posts and reshares from $2{,}819{,}338$
users for $4{,}663$ unique stories.
We filtered out stories posted or reshared more than $3{,}000$ times as well as stories whose number of posts or reshares taking place after the last decile of the observation period is greater than 1\%. Finally, we
filtered out fake stories at random until the percentage of fake stories is less than 15\%\footnote{\scriptsize We subsample the number of fake stories to resemble a more realistic scenario in which the number of fake
stories is \emph{small} in comparison with the total number of stories. We obtained qualitatively similar results without subsampling.}.
\begin{figure}[t]
      \centering
     \subfloat[Twitter]{\includegraphics[width=.5\textwidth]{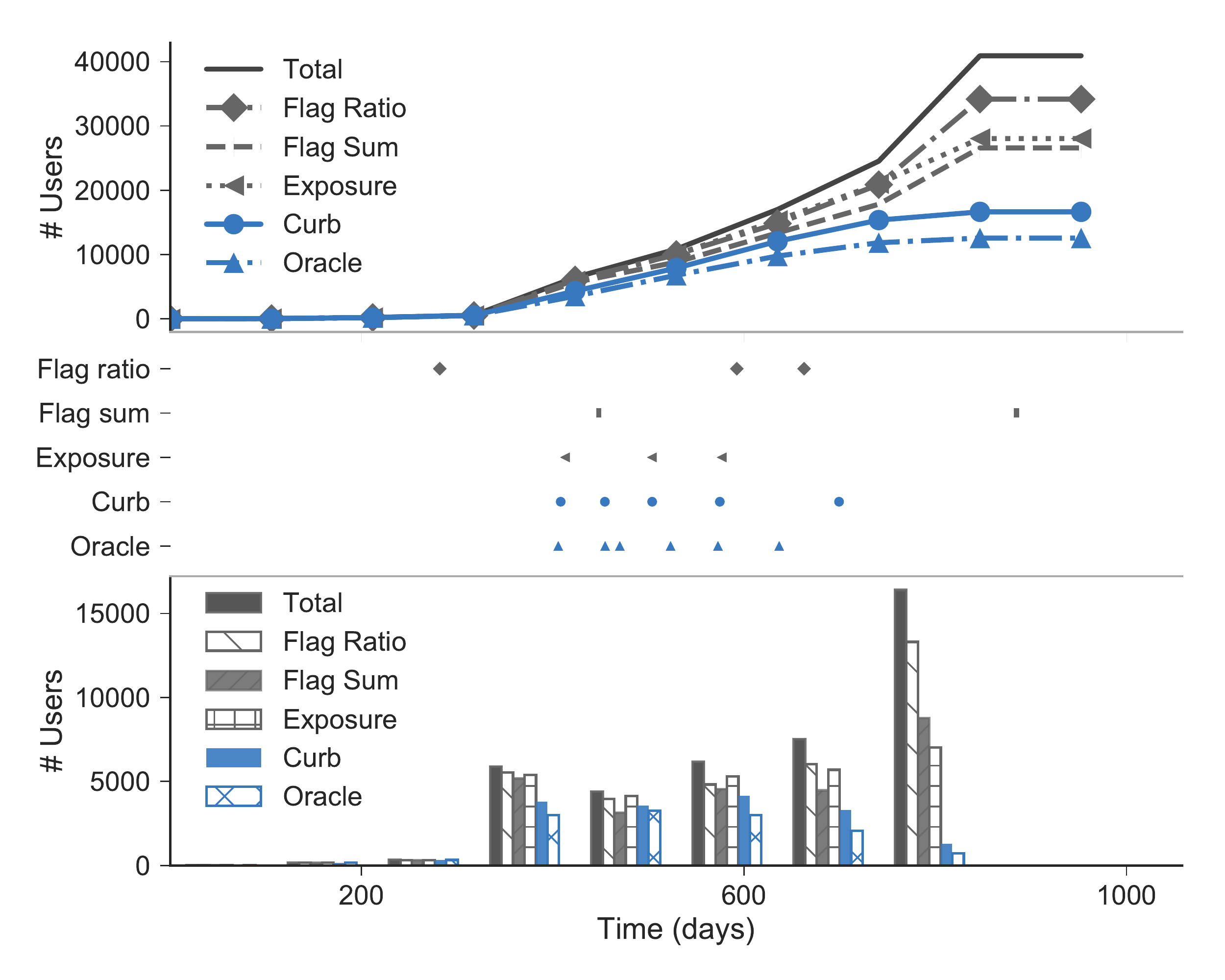}}
     \subfloat[Weibo]{\includegraphics[width=.5\textwidth]{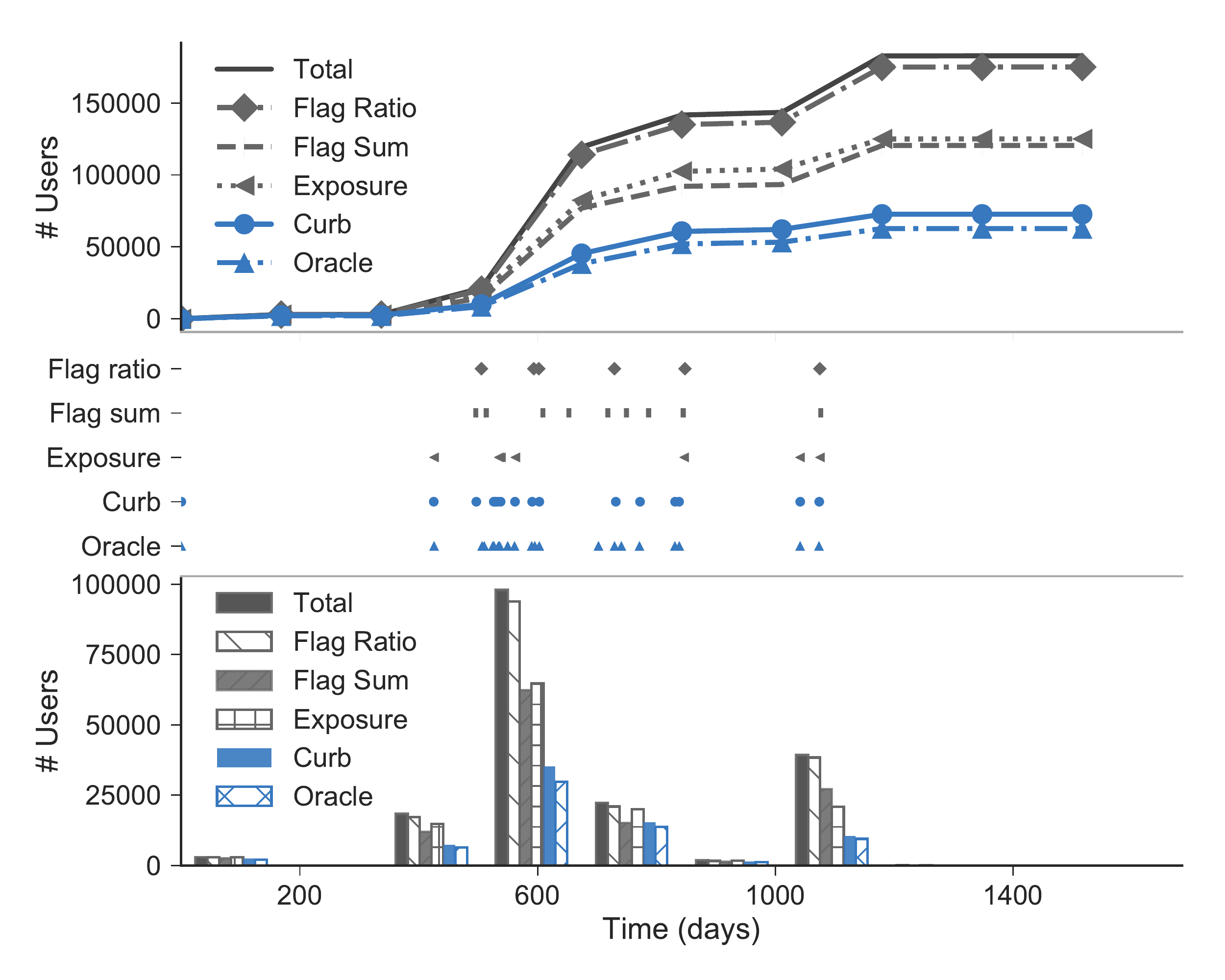}}
     \vspace{-3mm}
     \caption{Misinformation reduction vs. time.
     For both (a) and (b), the bottom panel shows the number of users exposed to misinformation at each time period, the middle panel shows when different methods send stories for fact checking, 
     and the top panel shows the cumulative number of users. The bars and lines show the different fact checking schedules and "Total" which 
     shows the numbers in absence of fact checking.
     Both \curb and Oracle are able to prevent the spread of misinformation before it becomes viral, whereas the other baselines cannot do so.
     The parameters for all methods are set so that the total number of fact-checked stories is $15 \pm 4$ for Twitter and $64 \pm 10$ for Weibo.
     %
     %
     %
      } \label{fig:time-reduction}
    \vspace{-4mm}
\end{figure}

%
%
%
After these preprocessing steps, our Twitter dataset consists of $28{,}486$ posts and reshares from $18{,}880$ users for $7$ fake stories and $39$ genuine stories and our Weibo dataset consists of $93{,}943$ posts and
reshares from $88{,}913$ users for $23$ fake stories and $133$ genuine stories.
Unfortunately, the datasets do not contain any information about the timing (or number) of exposures nor flags. To remedy the data unavailability, we generate user exposures based on user reshares and sample flags from Bernoulli distributions. We experiment with diffe\-rent parameter settings to ensure that our model's efficacy is robust throughout different settings. The details of the data generation steps are described in the following paragraph.
%

For each story, we sample exposure events using the exposure intensity defined by Eq.~\ref{eq:lambda-endogenous}, which does depend on the observed posts and reshares. There, we set $\gamma= 10^{-4}$ and $\omega = 10^{-5}$.
This choice of parameters results in approximately $10$ to $20$ exposures per post (or reshare) and a half-life of approximately $19$ hours per post (or reshare). Under this setting, our Twitter and Weibo datasets contain 313,357 and 1,054,449 exposures respectively.
%
%
Unless otherwise stated, for each exposure, we sample flags for fake and genuine stories using two bernoulli distributions with $p_{f = 1 | m = 1} = 0.3$ and $p_{f =1 | m = 0} = 0.01$, respectively.
Moreover, we estimate $p_d$ using the ratio between fake and overall number of stories, compute $p_{m | f = 1}$ and $p_{m | f = 0}$ from $p_{f = 1 | m}, p_{f = 0 | m}$, and $p_d$, and set
$\alpha/(\alpha + \beta) = (p_{f=1|d}  + p_{f = 0 | d}) p_d$.
Finally, note that we obtained qualitatively similar results for other experimental settings.

\vspace{-3mm}
\subsection{Evaluation metrics and baselines}
Throughout this section, we evaluate the performance via two metrics: precision and misinformation reduction. Precision is the fraction of fact checked stories that
are fake. Misinformation reduction is the fraction of unverified exposures that fact checking prevented.
%

We compare the performance of \curb against an ``Oracle", which is a variant of \curb and has access to the true flag probability $p_f$, and three baselines.
%
%
The first baseline (``Flag Ratio") samples the fact checking time for each story using an intensity which is proportional to the ratio between the number of flags and the total number of 
exposures, \ie, $u_s(t) = q_s ({\alpha + N^f_s(t)})/({\alpha + \beta + N^{e}_s(t)})$. 
The second baseline (``Flag Sum") sends a story for fact checking as soon as it accumulates a certain number of flags. 
The third baseline (``Exposure") samples the fact checking time for each story using an intensity which is proportional to the exposure intensity, \ie, $u_s(t) = q_s \lambda_s^e(t)$. 
Here, note that the Flag Sum baseline utilizes a deterministic policy while our method, the Oracle and the other baselines use a stochastic policy defined by an intensity 
function.
%
%
%
\begin{figure}[t]
      \centering
      \subfloat[Twitter]{\includegraphics[width=.3\textwidth]{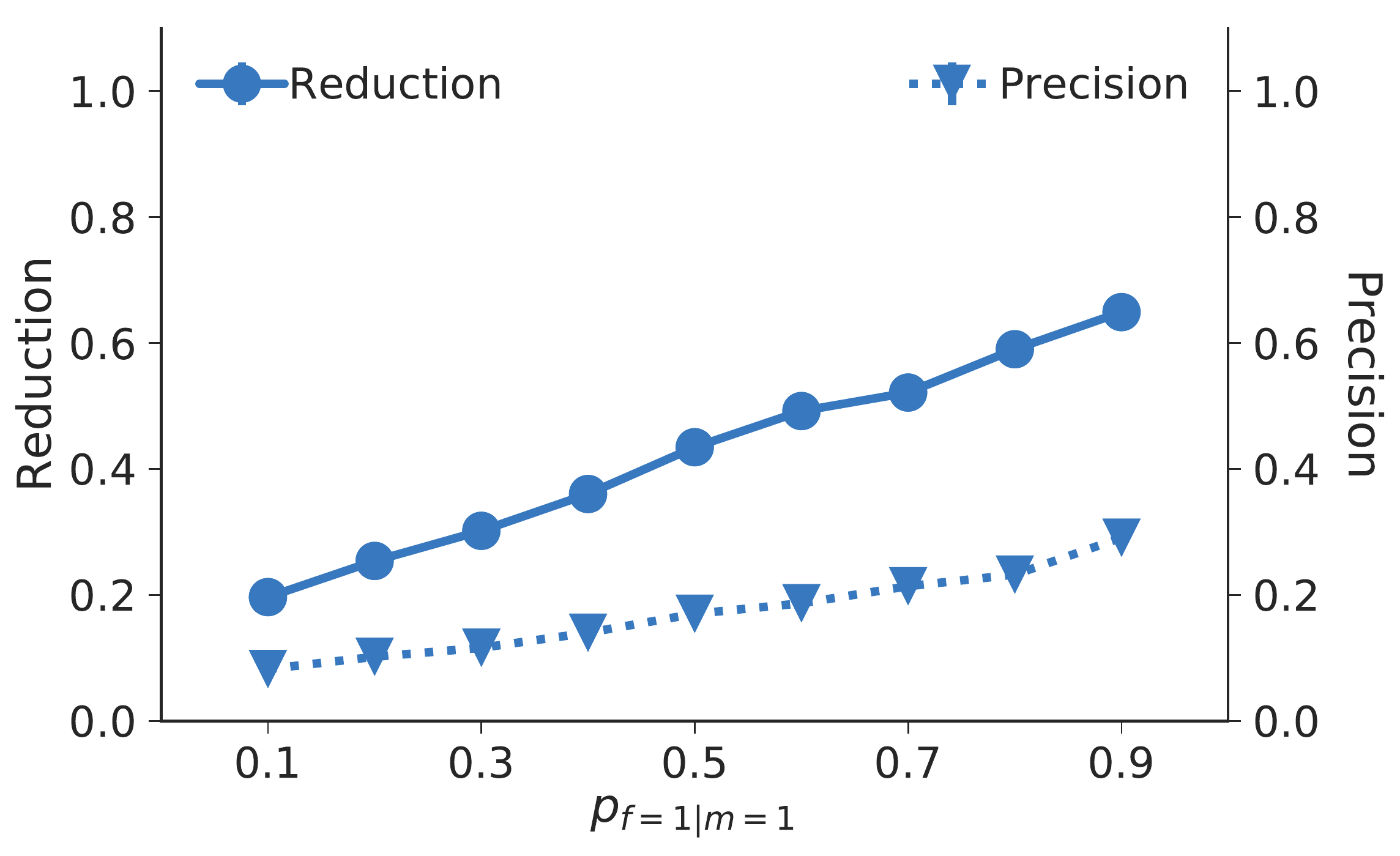}}  \hspace{4mm}
      \subfloat[Weibo]{\includegraphics[width=.3\textwidth]{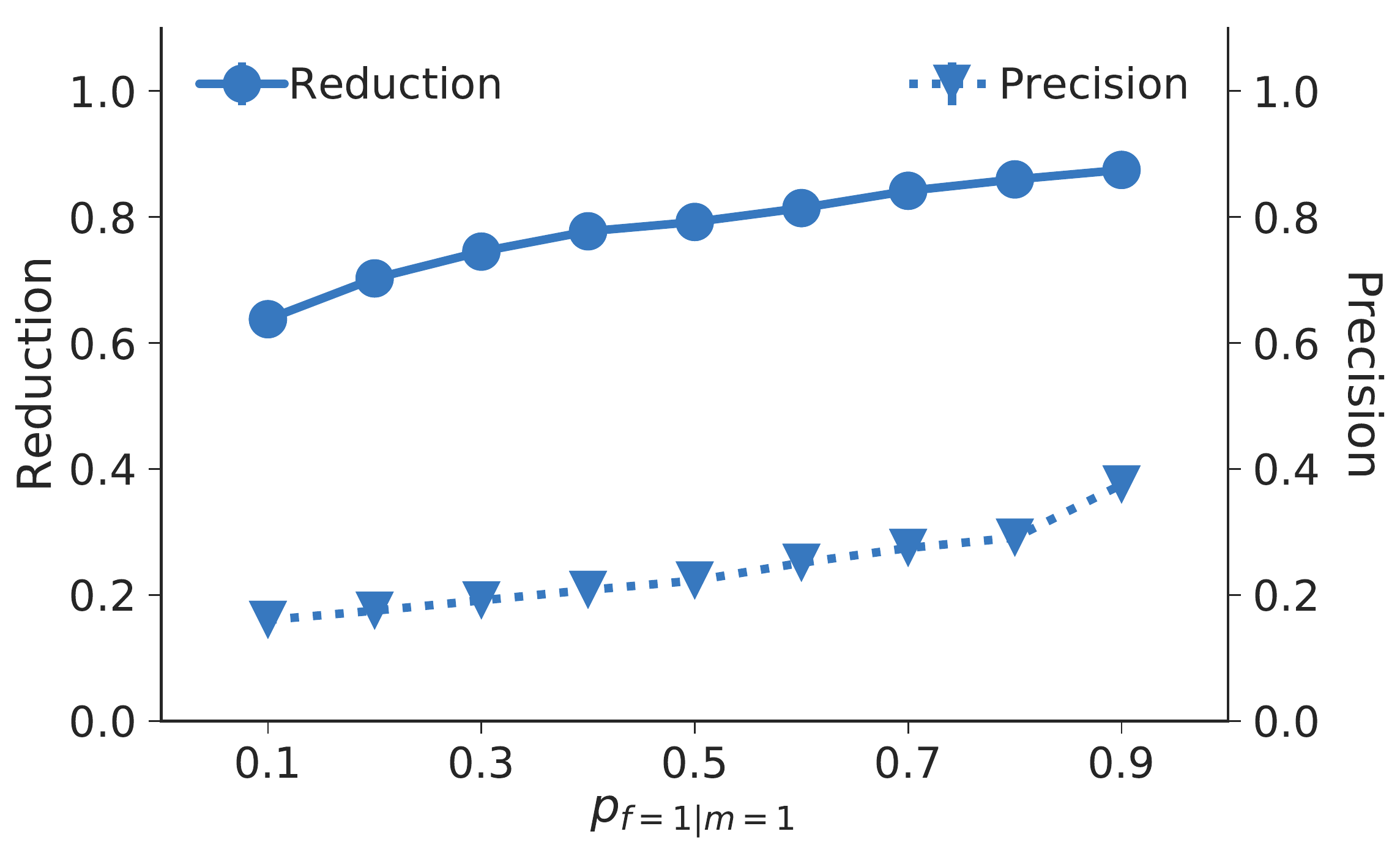}}
      \caption{Performance for different true positive rates, $p_{f = 1 | m = 1}$. Here, we set the false positive rate to $p_{f = 1 | m = 0}=0.01$.
      The higher the true positive rate, the \emph{better} the crowd is at spotting misinformation and the more effective \curb is.
      Total number of fact checked stories is $21 \pm 2$ for Twitter and $90 \pm 15$ for Weibo.
      } \label{fig:difficulty-number}
\end{figure}

\subsection{Solution quality}
%
%
%
We first evaluate the performance of our method, the oracle and the three baselines against the number of fact checking events. Figure~\ref{fig:tradeoff-number} 
summarizes the results, which show that:
(i) the more the number of fact checking events, the higher the average misinformation reduction for all methods; and,
(ii) Oracle and \curb achieve a comparable performance in both datasets, outperforming the baselines in most cases.
%

Next, we evaluate the performance in terms of misinformation reduction over time in Figure~\ref{fig:time-reduction}. We find that both \curb and the Oracle are able to prevent the spread 
of misinformation before it becomes viral. In contrast, the baselines are unable to spot misinformation that will become viral in time partly due to ignoring either the flags or the exposure 
dynamics.
%
%
%
%

\vspace{-3mm}
\subsection{Flagging behavior and misinformation evidence}
In this section, we first investigate how sensitive our algorithm is to the crowd'{}s flagging behavior and then explore the influence of the hyperparameters $\alpha$ and $\beta$
on the posterior estimate of the misinformation rate.

Intuitively, the more (less) accurate the crowd is at spotting misinformation, the more (less) effective our algorithm will be. Figure~\ref{fig:difficulty-number} confirms this intuition
by showing the performance of our algorithm for different true positive rates, $p_{f = 1 | m = 1}$.
%
%
%
%

\begin{figure*}[t]
      \centering
      \subfloat[Fake Story]{\includegraphics[width=.49\textwidth]{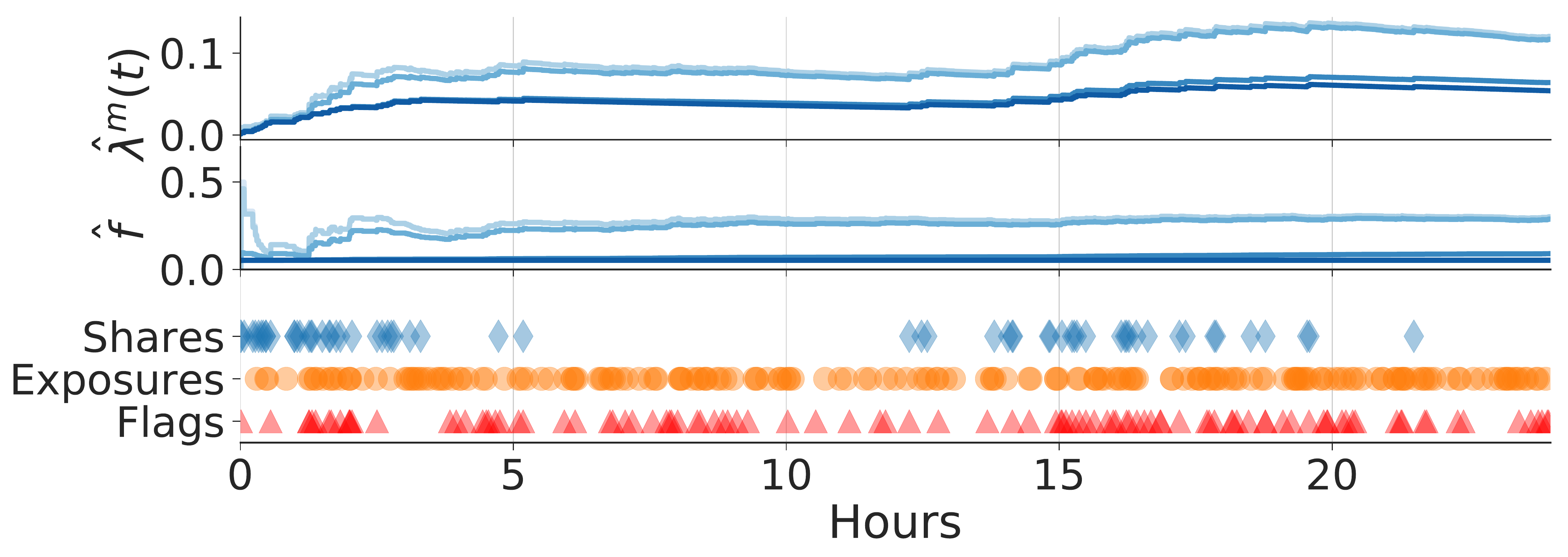}} \hspace{2mm}
      \subfloat[Genuine Story]{\includegraphics[width=.49\textwidth]{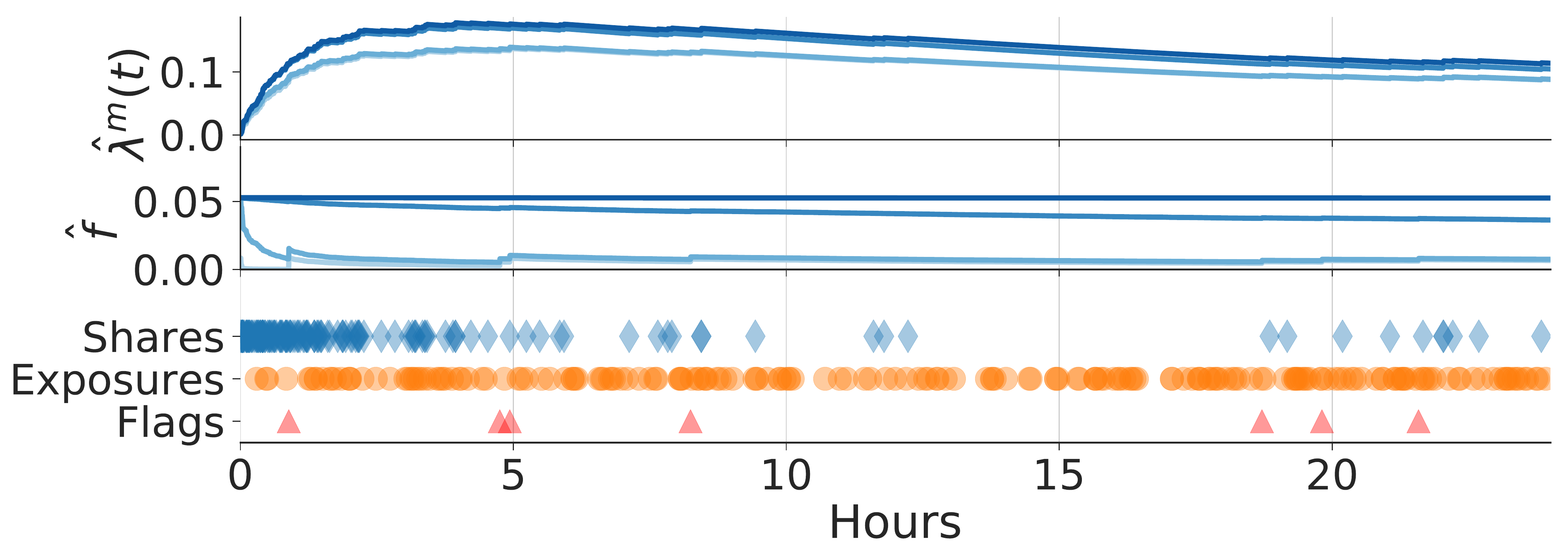}}
      \caption{Influence of the hyperparameter $\alpha$ on the posterior estimate of the misinformation rate, which the optimal intensity $u^{*}_s(t)$ depends on, and the posterior estimate of the
      flagging probability. In the top and middle panels, darker shades indicate stronger prior. The higher the value of $\alpha$, the smaller the effect of individual exposures and flags on the posterior estimate of both the misinformation rate and the flagging probability.}
      \label{fig:hyperparameters}
\end{figure*}

Next, we investigate the influence of the hyperparameter $\alpha$ on the posterior estimate of the misinformation rate $\lambda_s^m(t)$, which the optimal intensity $u^{*}(t)$ depends on,
and the posterior estimate of the flagging probability $\EE[f_s | N_s^e(t), N_s^f(t)] = \hat{f}_s$ by examining an example of a genuine and a fake story.
%
%
Figure~\ref{fig:hyperparameters} summarizes the results, which show that:
(i) the higher the value of $\alpha$, the smaller the effect of individual exposures and flags on the posterior estimate of both the misinformation rate and the flagging probability when
a story starts spreading---the lower the variance of the estimates over time; and,
(ii) the posterior estimate of the misinformation rate, which the optimal intensity $u^{*}(t)$ depend on, changes in a more intricate way than the posterior estimate of the flagging probability since
it combines the latter with the exposure intensity. This enables our algorithm to be more cautious with viral stories even if the evidence of misinformation is weak.
%
%

%% file: 050conclusions.tex
In this paper, we have introduced an efficient online algorithm, \curb, that leverages the crowd to detect and prevent the spread 
of misinformation in online social networking sites. 
In doing so, we establish an unexplored connection between stochastic online optimal control of stochastic differential equations (SDEs) 
with jumps, survival analysis, and Bayesian inference.
We experimented with two real-world datasets gathered from Twitter and Weibo and showed that our algorithm may be able to effectively 
reduce the spread of misinformation. 

There are many interesting directions for future work.
For example, we assumed every person in the crowd is equally good (or bad) at flagging misinformation. It would be interesting to relax this assumption, infer 
each person'{}s trustworthiness, and design algorithms that are robust to adversarial behavior from part of the crowd.
Moreover, we considered that stories are independent and the probability that a story is misinformation given that a user did (not) flag a story is equal for all stories. 
However, stories may be dependent and the probability that a story is misinformation given a user did (not) flag it may be different for stories supported by different 
sources (or domains).
%
%
Finally, our algorithm optimizes a quadratic loss of the posterior estimate of the misinformation rate, however, it would be useful to derive fact checking intensities 
for other losses capturing more nuanced goals.
%

%% file: 060appendix.tex
\begin{lemma} \label{lem:dJ}
Let $x(t)$ be a jump-diffusion process defined by the following SDE:
\begin{equation*}
dx(t) = f(x(t), t) dt + g(x(t), t) z(t) dN^{e}(t) + h(x(t), t) (1-M(t)) dN^p(t) + k(x(t), t) dM(t),
\end{equation*}
where $M(t)$, $N^{e}(t)$, $N^p(t)$ are independent jump processes, $N_f(t)$ is a jump process whose increment is defined as 
$dN_f(t) = w(t) dN^{e}(t),\, w(t) \in \{0, 1\}$, and $z(t) \in \{ 0, 1\}$. If the function $F(M(t), N^{e}(t), N^f(t), N^p(t), x(t), t)$ is once continuously 
differentiable in $x(t)$ and $t$, then, 
\begin{align*}
  &dF (M(t), N(t), N^f(t), N^p(t), x(t), t) \\
    &= \left[     F  (  M, N^{e} + 1, N^{f} + 1, N^p, x + g,  t ) w(t)z(t) + F (   M, N^{e} + 1, N^{f} + 1 , N^p, x,  t ) w(t)(1 - z(t))  \right.    \\
    &\quad+  F (   M, N^{e} + 1, N^{f}, N^p, x + g , t ) (1 - w(t))(z(t))  +  F (   M, N^{e} + 1, N^f , N^p, x, t  ) (1 - w(t))(1 - z(t)) \\ 
    &\quad\left. - F(M, N^{e}, N^f, N^p, x, t)  \right] dN^{e}(t) + \left[ F (   M, N^{e}, N^f , N^p+1, x + h, t  ) \right. \\ 
    &\quad \left. - F( M, N^{e}, N^f, N^p, x, t)   \right]  dN^p(t) + \left[F(M+1, N^{e}, N^f, N^p, x+k, t) \right. \\
    &\quad \left. - F(M, N^{e}, N^f, N^p, x, t) \right] dM(t) + F_t dt + f F_x  dt, 
  \end{align*}
where for notational simplicity we dropped the arguments of the functions f, g, h, k.
\end{lemma}
\begin{xsketch}
The differential of $J$ can be found using Ito'{}s calculus~\cite{hanson2007applied}. 
In particular, using that the bilinear forms $dt dN^e(t) = dt dN^p(t) = dt dM(t) = 0$ and $dN^e(t) dM(t) = 0$ and $N^p(t) dM(t) = 0$ by the zero-one jump law.
\end{xsketch}

%